\documentclass[aps,preprint,epsfig,showpacs]{revtex4}
\usepackage{graphicx}
\usepackage{epsfig}
\usepackage{times}
\newcommand{\be}{\begin{equation}}
\newcommand{\ee}{\end{equation}}
\newcommand{\bea}{\begin{eqnarray}}
\newcommand{\eea}{\end{eqnarray}}

\begin{document}
\title{Effects of Molecular Crowding on 
stretching of polymers in poor solvent}
\author{Amit Raj Singh$^\star,1$, Debaprasad Giri${^1}$, Sanjay Kumar$^1$}
\affiliation{${^1}$Department of Physics, Banaras Hindu University,
Varanasi 221 005, India \\
${^1}$ Department of Applied Physics, Institute of Technology, \\
Banaras Hindu University, Varanasi 221 005, India}
\email{amitraj_phy@rediffmail.com}
\date{\today}

\begin{abstract}
We consider a linear polymer chain in a disordered environment
modeled by percolation clusters on a square lattice. The disordered
environment is meant to roughly represent molecular crowding as 
seen in cells.  The model may be viewed as the simplest
representation of biopolymers in a cell.
We show the existence of intermediate states during stretching
arising as a consequence of  molecular crowding.
In the constant distance ensemble the force-extension curves exhibit 
oscillations. We observe the emergence of two or more peaks
in the probability distribution curves signaling the coexistence of
different states and indicating that the transition is
discontinuous unlike what is observed in the absence of molecular crowding.

\end{abstract}
%\pacs{64.60.ah, 87.10Hk, 82.37Rs, 87.15Cc}
\pacs{82.35.Jk, 36.20.Ey, 64.90.+b}
%\pacs{82.37.Rs, 36.20.Ey, 64.60.ah}
\maketitle

\section{introduction}
Every major change in cellular systems involves mechanical movement 
at a single molecule level. Recent advances in single molecule force 
spectroscopy (SMFS) {\it e.g} optical tweezers, atomic force 
microscope, {\it etc.} have allowed cellular processes to be examined 
at the single molecule level \cite{rief1,rief2,busta1, busta2}.   
Moreover, by applying a force of the order of pN on an isolated protein 
in vitro, the  response of the force has been studied in order to 
understand the elastic, structural and functional properties of 
proteins \cite{rief2,busta3,lemak}. Modeling of proteins using 
simplified interactions amenable to statistical mechanics analysis
have been used extensively to theoretically understand the outcomes 
of these experiments \cite{fixman,degenes,doi,vander}.  
Among the theoretical approaches lattice models despite their 
simplicity, have proved to be quite predictive and have provided 
much important information about cellular 
processes \cite{li0,somen1,maren1,maren2,kumar}.

New challenges have arisen in the area of protein folding when 
removed from an artificial, controlled environment in vitro and 
relocated to the cellular environment \cite{ellis, matou,goodsell,mueller}. 
Cells have a very crowded environment because they are composed of 
many different kinds of biomolecules that may occupy a large fraction 
($\approx 40 \%$) of the total volume (Fig. 1). This condition leads to a 
phenomenon called ``volume exclusion" which is caused by the steric 
repulsion between different molecules \cite{minton,cheung}. 
It is now known that molecular crowding can influence the stability, 
dynamics and function of proteins. Thus far, in theoretical modeling, 
the cellular environment has been considered as homogeneous with 
each monomer (amino acid) interacting with all its nearest 
neighbors \cite{maren2,kumar}, of which there is a fixed number for 
each given lattice. In vivo, the cellular environment  and the interactions 
involved in the stability of proteins are no longer homogeneous. It is 
essential to somehow model this disordered media say via the introduction 
of randomness into the connectivity of the underlying lattice. 
Linear polymer chains trapped in a porous (random) media have been 
studied in detail because of the technological importance in filtration, 
gel permeation chromatography etc. \cite{bkc,bkc1,janke}. However, 
the response of a polymer to a force remains an elusive problem which 
has thus far not been studied in detail.

\begin{figure}[t]
\centerline{\includegraphics[width=3.5in]{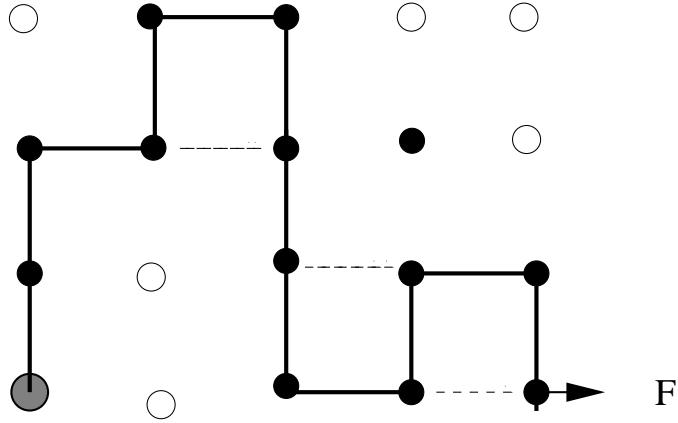}}
%\vspace{-1.5in}
\caption{Schematic representation of a polymer chain in a disordered 
medium. The black and white circles represent available and unavailable 
sites respectively. The imposed restriction of unavailability of 
certain sites  gives rise to a ``volume exclusion" of about $40 \%$ 
as seen in the cell. One end of the polymer chain is kept fixed 
while a force $F$ is applied at the other end. Dashed line 
corresponds to nearest neighbor attractive interaction among monomers.
}
\label{fig-1}
\end{figure}

The aim of this paper is to study the effect of an applied force
on a polymer in an artificially reproduced environment with molecular 
crowding similar to what is observed in a cell. In general the effect 
of the force on the reaction co-ordinate (end-to-end distance) \cite{bustar4} 
is mainly determined by the competition between a loss of configurational 
entropy and a gain in internal energy due to the stretching of the protein 
caused by the applied force.
The confinement of the proteins to a restricted portion of a cell leads 
to a further loss in entropy because of the molecular crowding and this may
affect the behavior of the proteins.
Since SMFS experiments are performed on proteins with only a few monomers
we expect that modeling the effect of a force on a {\it finite} chain in 
a disordered environment may provide a better understanding of the 
unfolding process in vivo.  In disordered media information 
about the dependence of the reaction co-ordinate and the probability 
distribution of the reaction co-ordinate on the parameters of the system 
is difficult to obtain analytically and one therefore has to resort to 
numerical studies. 

\begin{figure*}[t]
\vspace {.2in}
\centerline{\includegraphics[width=6.5in]{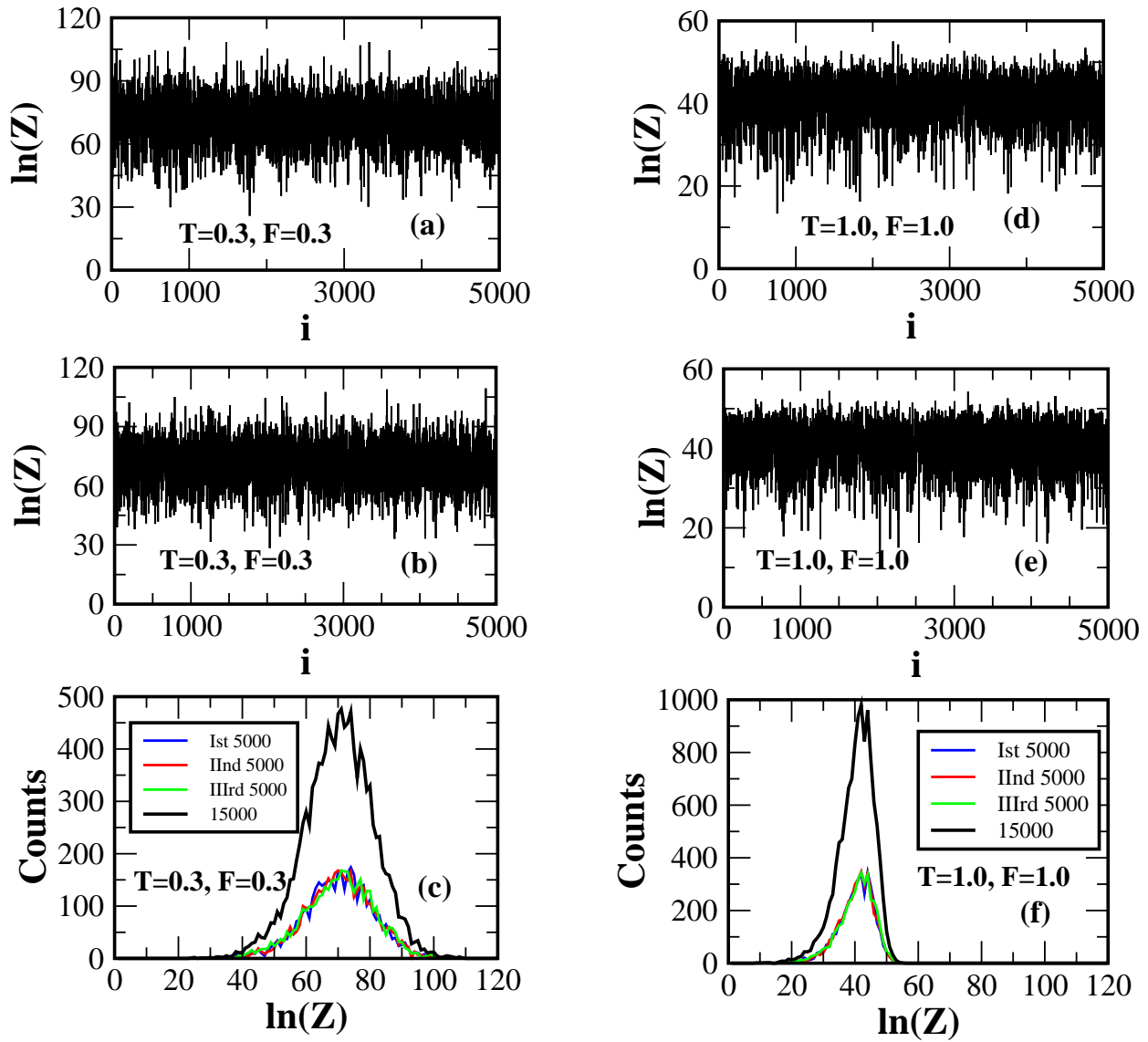}}
%\vspace{-1.5in}
\caption{(Color online) Fig. (a) and (b)  show the variation 
of $\ln Z$ for two different sets of realizations (5000 each) 
at $T=0.3$ and $F=0.3$. 
In Fig. (c) we plot a histogram of $\ln Z$. Note that all 
the three histograms overlap within statistical error bar.  
Figures (d-f) are for $T=1.0$ and $F=1.0$.
}
\label{fig-2}
\end{figure*}

The freely jointed chain (FJC) and the worm like chain (WLC) 
(developed for polymers) \cite{doi,vander}
are models which have been used extensively in order to understand
the unfolding process and they correctly describe the force 
extension ($F-x$) curves in the intermediate and high force regime. 
However, these models ignore excluded volume effects in their 
description. Off lattice simulations do provide much important 
information about unfolding processes \cite{li0}, but the exact 
density of states over the entire force regime are difficult
to obtain. Recently, exact and complete information about the 
density of states was obtained using exact enumeration technique. 
Such studies can reproduce many qualitative features 
of the force induced transitions as well as reveal many new 
insights into the mechanisms governing the cellular 
processes \cite{somen1,maren1,maren2,gk,sk,kjjg}. 

\section{Model}

We model the polymer chain as self-attracting self-avoiding 
walks (SASAWs) whose one end is kept fixed and a force is applied at 
the other end. The confinement imposed by molecular crowding has 
been modeled by generating site percolation 
clusters \cite{bkc1, janke, stauffer} on the underlying square 
lattice of $M$ sites with a fixed density $p$ of available 
sites.
In order to model the `` volume exclusion", which is about $40\%$ of the
total volume \cite{minton,cheung}, we choose a density $p$ of 
available sites just above the percolation threshold 
$p_c (= 0.5928)$ \cite{janke}. It means that a SASAW
cannot access about $40\%$ of the volume as
happens in the cell. The percolation process gives rise to a
distribution of clusters, which will vary in size from isolated 
disorder sites upto clusters with an extent spanning the entire system.
Due to the randomness of the clusters the ground state of the system 
will not be fixed (as is the case on a regular lattice), but may 
vary with each realization.
Because of the steric repulsion between the molecule of interest and the
surrounding biomolecules \cite{minton,cheung}, in the present model we
do not consider interactions between the polymer and the disorder sites.
The model can readily be extended to three dimensions (3D) \cite{maren2}. 
The qualitative nature of the $F-x$ curve should remain the same.
But in the case of 3D lattices the density of states can be calculated 
exactly for walks of length  up to only 20 steps and one therefore
has to use other methods {\it e.g.} Monte Carlo simulation \cite{maren2},
molecular dynamics \cite{li0} etc. to study such systems.

Since molecular crowding is a dynamical phenomenon
involving the appearance, disappearance and movement of voids,
the internal structure of the cell changes continuously.
To model this  we perform an averaging over many
realizations while keeping $p_c$ constant.
This ensures that the concentration of the
crowding agent does not change, but the internal structure may change
as happens in the cell. Since we are near to percolation threshold,
every disorder realization constitutes the partition function which 
is non zero. 
The partition function of the $i^{th}$ realization of the disorder
configuration may be written as

\begin{equation}
Z_i = \sum_{(N_p, |x|)} C_N^{(i)} (N_p, | x |)
u^{N_p} \omega^{| x |}.
\end{equation}
Here $C_N^{(i)} (N_p, | x |)$ is
the number of distinct conformations of walks of length $N$ in the $i$-th 
realization of the disordered with $N_p$ 
non-bonded nearest neighbor pairs and whose end-points are a 
distance $x$ apart.  $\omega$ is the Boltzmann weight for the
force defined as $\exp[\beta (\vec{F} \cdot \hat{x})]$, where $\hat{x}$
is the unit vector along the $x$-axis. $\beta$ is defined as
$\frac{1}{k T}$ where $k$ is the Boltzmann constant and $T$ is the
temperature. $u = \exp (-\beta \epsilon)$ is the Boltzmann weight  
of nearest neighbor interactions with energy $\epsilon$. 
In the following, we set $\epsilon /k = 1$ and  focus 
our discussion on the force induced globule-coil transition on the 
percolative lattice. Throughout this paper $[...]$ denotes an average 
over various realizations and $\langle...\rangle $ denotes thermal 
averaging. In the present study we enumerate all possible walks of 
length $N= 45$ steps on given percolation clusters using  
15000 realizations  ($N_{tot}$) of the percolation process.
For the sake of comparison we also enumerate walks of the same 
length on the square lattice with no disorder.  
It has been shown in previous studies that the chain length considered here 
is sufficient to predict the correct qualitative behavior and while
increasing the chain length yields better estimates of say the phase 
boundary  the qualitative features of the phase-diagram remain the 
same\cite{sk,kjjg}.

The limit $T \rightarrow \infty$ corresponds to pure 
self-avoiding walks (SAWs) {\it i.e.} polymers in a good 
solvent \cite{degenes}. 
We reproduced the scaling proposed  by Blavatska and Janke \cite{janke} 
for SAWs 
on a percolation cluster as well as for the square lattice. 
In Figs. 2(a) and (b) we plot $\ln Z_i$ with $i$ for two 
different sets (5000 each) of realizations. In Fig. 2(c), we have 
plotted the histogram of $\ln Z$. The resulting distributions overlap 
(within statistical error bar) with each other having a common peak.

\section{Results}

In studying properties of disordered systems one encounters two types
of averages in the literature namely the annealed average and the quenched
average \cite{bkc,li}. In the case of annealed averaging conformational
changes in the disorder have time scales which are very fast compared
to the motion of the polymers. Therefore, in the annealed case, averaging
has to be done over all possible conformations of the disorder implying
that $N_{tot} \rightarrow \infty$. In the case of quenched disorder the
biopolymers unfold very fast compared to the conformational changes 
(if any) of the disorder sites. 
In fact neither type of averaging is  entirely appropriate as
far as the unfolding of a biopolymer in the presence of continuously
moving crowding agents  is concerned. It may be noted that unfolding takes 
place on time-scales on the order of micro seconds to  a few seconds.
In such a relatively short time span the internal structure of a cell 
can change substantially but not enough to access
all possible conformations and hence increasing the number of 
realization toward the limit of infinity is not an experimental 
prerequisite. In view of the time scale involved in unfolding, 
we restrict ourselves to averaging over a finite but
large set of realizations which may roughly mimic a real system 
in vivo. The number of realizations considered here is almost 
twenty times more than  previous studies \cite{bkc1}.
The approximate annealed average of the reaction co-ordinate 
in this case may be defined as 

\begin{equation}
\langle [x]\rangle_D = \frac{\frac{1}{N_{tot}} \sum_i \sum_{(N_p, | x |)} 
x^{(i)} C_N^{(i)}(N_p, | x |)u^{N_p} \omega^{| x |}}{Z}
\end{equation}

where $ Z = \sum_i Z_i/N_{tot}$ and the summation is over $N_{tot}$ 
realizations of the disorder.  It may be noted that in the limit 
$N_{tot} \rightarrow \infty$ the annealed average and
the pure system will give the same results and the two summations 
of Eq. (2) can be interchanged. In this case there are roughly 
$2^{M}$ possible conformations and hence summation over all of them 
is computationally impossible.  The suffix ``D" in Eq. (2) 
corresponds to an approximate annealed averaging that depends on 
the  disorder realization. 
%Therefore, we call it an approximate 
%annealed average. 
The quenched average of the reaction co-ordinate for the $i$-th realization 
is given by
\begin{equation}
\langle x^{(i)}\rangle_Q = \frac{\sum_{(N_p, | x |)} x^{(i)} C_N^{(i)} 
(N_p, | x |) u^{N_p} \omega^{| x |}}{Z_i} 
\end{equation}
The sample average over Eq. (3) can be written as 
\begin{equation}
[\langle x\rangle_Q] = \sum_i \langle x^{(i)}\rangle_Q/N_{tot} 
\end{equation}
which we call an average over quenched disorder.

\begin{figure*}[t]
\vspace {.2in}
\centerline{\includegraphics[width=6.5in]{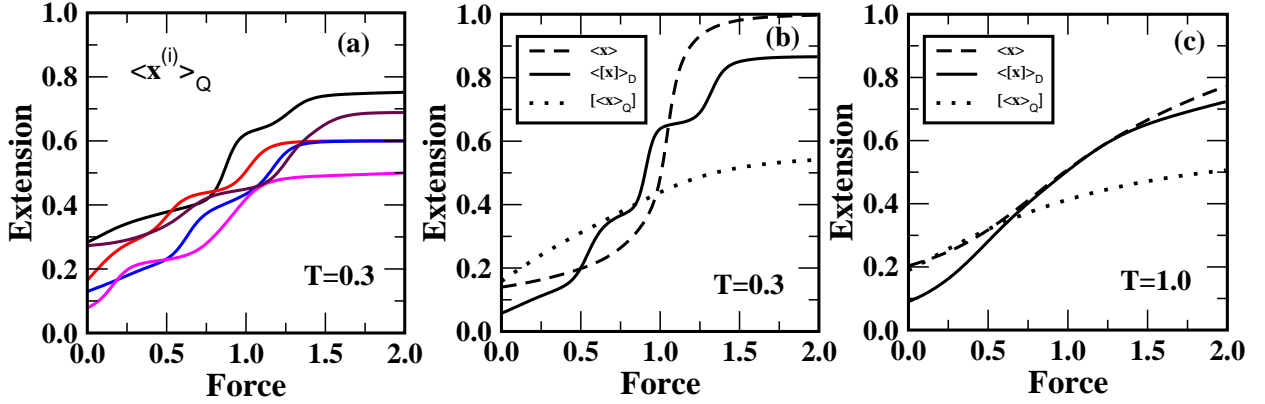}}
%\vspace{-1.5in}
\caption{(Color online) Figures show the variation of extension with force. 
In Figure 3(a) we show some of the representative F-x curves 
for different disorder realizations at $T = 0.3$. Figure 3(b) 
and (c) show the approximate annealed average and sample averaged 
quenched disorder at low $T$ (0.3) and high $T$ (1.0). In these 
plots, we also show the pure case for the comparison. The individual
realization and approximate annealed average exhibit multistep 
plateaus which are absent in the sample 
averaged quenched and pure cases.}
\label{fig-3}
\end{figure*}

\subsection{Analysis in constant force ensemble}

In Fig. 3(a) we show some of the representative plots of 
quenched averaged reaction 
coordinate obtained from Eq. (3) for different realizations. In Fig. 3(b) we
plot the extension versus the force for the approximate 
annealed disorder average ($\langle [x]\rangle_D$) and the sample averaged 
quenched disorder ($[\langle x\rangle_Q]$) at low temperature ($T = 0.3$).
For the sake of completeness we also plot $\langle x \rangle$ versus 
$F$ for the pure case at the same temperature. 
In the approximate annealed case we find  multistep plateaus 
which are absent in the pure case. 
Such plateaus have been observed in the pure case at much 
lower $T$ during force induced
unfolding \cite{maren2,kjjg}. It appears that disorder reduces the 
entropy of the system (making the solvent poorer) which causes 
the emergence of such plateaus. We expect qualitatively similar 
behavior for other sets of realizations with a finite $N_{tot}$ and 
these plateaus will vanish and approach the pure case in the 
limit $N_{tot} \rightarrow \infty$. 
Notably every specific realization shows such plateaus, which 
are induced by the disorder. 
However sample averaging [Eq. (4)] over quenched disorder [Eq. (3)], 
smoothen  the plateaus. At high temperature the entropy of the 
system is high enough that any effects of disorder vanish. As a consequence 
of this the force-extension curve overlaps with the pure case as 
can be seen from Fig. 3(c). 

\begin{figure*}[t]
\vspace{0.6in}
\centerline{\includegraphics[width=6.5in]{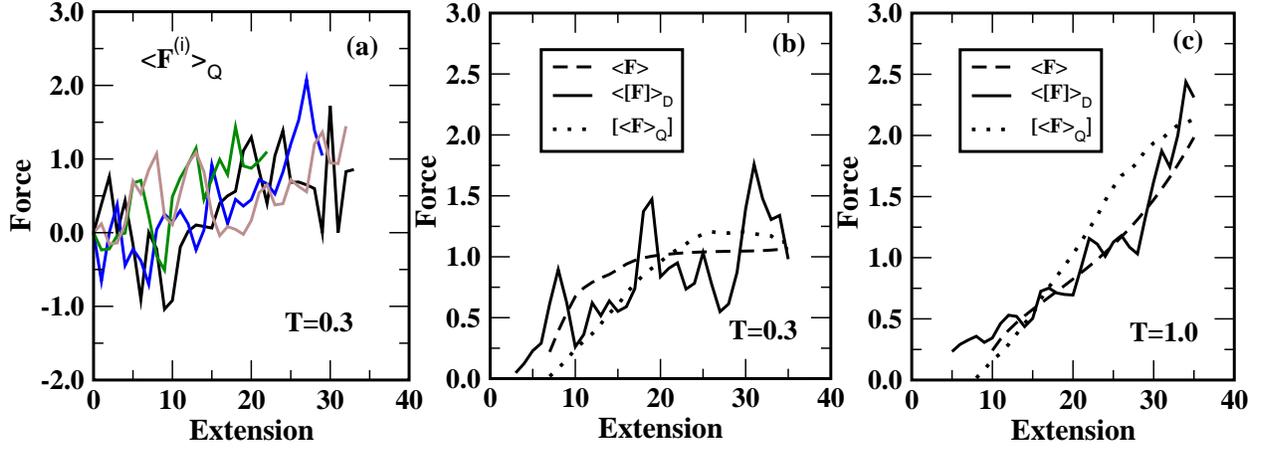}}
\caption{(Color online) Same as Figure 3, but at constant distance ensemble. 
The strong oscillations can be seen here for individual disorder
realizations and in the approximate annealed case, but absent in 
the pure and sample averaged quenched cases. 
}
\label{fig-4}
\end{figure*}

\subsection{Analysis in constant distance ensemble}

It is known that in  protein folding, the protein acquires a
unique (native) conformation among a large number of conformations
and many biological functions depend on this.  However, in a
``disordered protein" \cite{romero, tompa} there is no unique
native conformation and diseases such as Alzheimer are the result
of that. In fact the ground state entropy of such molecules
is comparatively larger than the native conformations (unique
in the case of a protein) but much less than the entropy associated
with the globule of the same length \cite{romero}.
It is therefore important to have a reference calculation which
predicts the shape of these curves in quasi-equilibrium in order
to precisely estimate kinetic effects in the experiments or in
molecular or Langevin dynamics simulations [13].
In the case of polymers we find that the residual entropy of the
globule on the percolation cluster is much less than the entropy
of the globule (pure case) on the square lattice.
Moreover, on the percolation cluster, the number of interactions per
monomer is not the same as on a pure square lattice and this
induces a heterogeneity in interactions along the chain even
for a homopolymer. The number of nearest neighbors in this
case varies between 0 and 2 in comparison to the pure lattice
where this number is always 2. Because of this the present
model may give some intrinsic features close to such molecules
because of the induced heterogeneity in the interaction. 
Since atomic force microscopy (AFM) works in the constant 
distance ensemble (CDE) \cite{gk,kjjg},
we also calculate $\langle F\rangle$ in the CDE and plotted it 
against $x$ (Fig. 4). It is interesting to note that the model 
presented here shows oscillations in the $F-x$ curve for the 
approximate annealed case. However, for the sample averaged 
quenched disorder and pure cases such oscillations are absent.

\subsection{Probability Distribution}

It has been shown that the probability distribution curves 
$P(x)$  provide important information about cellular 
processes \cite{gk,sk}. $P(x)$ can be calculated from the following 
expressions for the approximate annealed and averaged quenched cases:
\begin{equation}
P_{A}(x)=\frac{\sum_i \sum_{N_P} C_N^{(i)} (N_p, | x |) u^{N_p} 
\omega^{| x |}}{\sum_i Z_i}
\end{equation}
 and 
\begin{equation}
P_{Q}(x)=\frac{1}{N_{tot}}\sum_i [\frac{\sum_{N_P} 
C_N^{(i)} (N_p, | x |) u^{N_p} \omega^{| x |}}{Z_i}]
\end{equation}

\begin{figure*}[t]
\vspace{.5in}
\centerline{\includegraphics[width=6.5in]{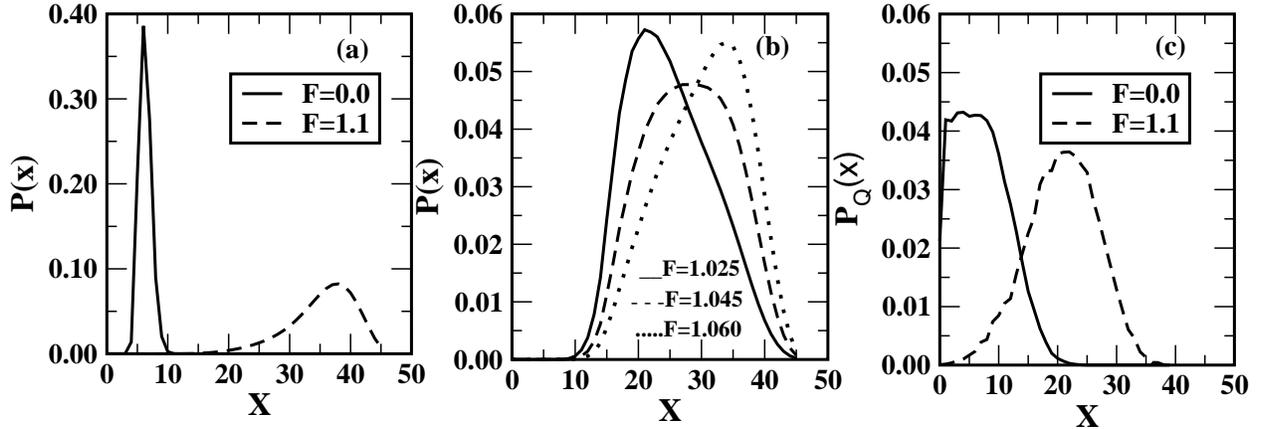}}
\caption{Figures (a - c) show the probability distribution curves 
for the pure and sample averaged quenched cases at $T=0.3$ at 
different forces.
(a) For $F = 0$ (pure case), the peak position corresponds 
to the folded state, while at $F=1.1$ the peak position 
corresponds to the stretched state; (b) Near the transition 
temperature the peak broadens indicating the transition 
is continuous. Figure (c) is for the sample averaged 
quenched disorder. One can see the effect of disorder induced 
entropy on the sample averaged quenched disorder which shows 
the broadening of the peaks at $F=0$ and $F=1.1$.
}
\label{fig-5}
\end{figure*}
For the pure, approximate annealed and sample averaged quenched 
disorder cases, the probability distribution 
curves shown in Figs. 5 and 6 display qualitatively similar behavior at 
low and high forces with $T=0.3$. For $F = 0$ we find a peak position 
corresponding to the collapsed state. At high  $F$ it peaks 
around the stretched state. 
The collapse transition is of second order in two dimensions (2D) \cite{zhou}
which can be seen from the probability distribution curve 
shown in Fig. 5(b) for the pure case.  
Near the transition point the peak broadens, which indicates that the 
transition is continuous (Fig. 5(b)). 
However, in the approximate annealed case [Figs. 6(a)-(l)], 
one can see the emergence and 
disappearance of  peaks for different $F$ at $T = 0.3$. 
With increasing $F$ we see that the height of one  peak increases 
while others decrease. 
We find at many different forces  that the height of two peaks (at different
positions) becomes equal indicating the coexistence of two states. These 
features are observed for each and every realization of disorder.  
This gives the signature that the transition is no longer
continuous.

\section{Conclusions}

Recently, Yuan {\it et al.} \cite{yuan} studied the effect of 
concentration of dextran (disorder sites) which is quite below 
the $p_c$ on the mechanical stability of protein molecules. They
found that average force increases with concentration in presence 
of crowding agents (dextran).  For the low 
concentration, force increases linearly, but above than $30\%$ 
concentration it is no more linear. Our early calculation for 
small realization do show such behavior and we find that 
there is a crossover when concentration approaches 
to $p_c$ \cite{amit}.

\begin{figure*}[t]
\centerline{\includegraphics[width=6.5in]{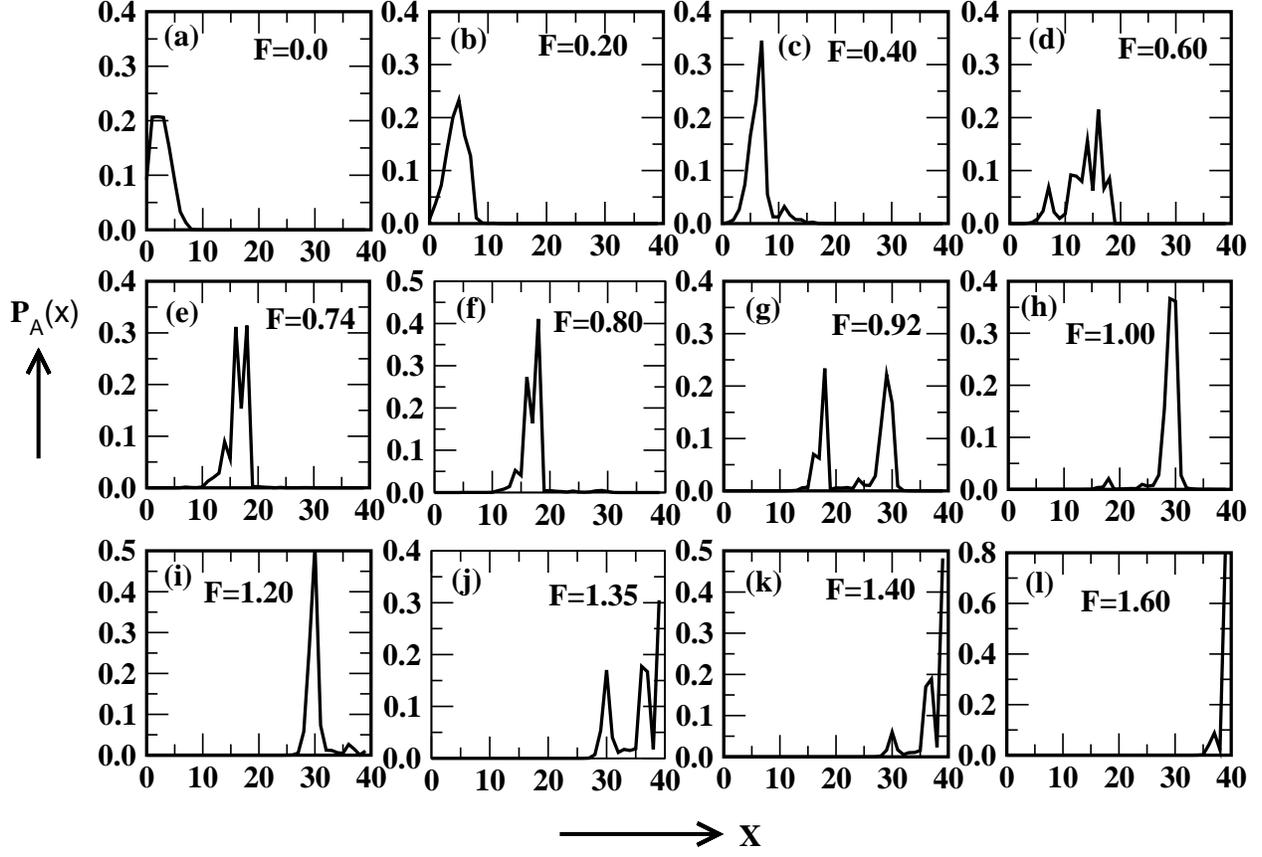}}
\caption{Figures (a - l) show the variation in the probability 
distribution curves for the approximate annealed case for
different forces at fixed temperature $T=0.3$. The appearance 
of peaks at different $F$ shows the existence of intermediate 
states induced by the force. The coexistence of peaks at 
different $F$ gives a clear signature that the transition 
is no longer continuous. 
}
\label{fig-6}
\end{figure*}
In this work we have reproduced most of the studied behavior 
of a polymer chain at force $F= 0$ {\it e. g.} scaling of the 
probability distribution curve (SAWs) for pure and disordered 
lattices \cite{janke}. Our results are also consistent with 
earlier studies, { \it i.e} at $F = 0$, $x$ is less than the 
pure case for approximate annealed averaging while greater for 
sample averaged quenched case (Fig. 3(b)) \cite{bkc,bkc1}. 
The approximate annealed 
average in fact represents the typical effect of the quenched disorder
case. However, entropy induced by sample averaging over quenched 
realization smoothens the reaction coordinate and we therefore
observe a monotonic increasing effect in the force-extension curve.
Similar features have also been seen for individual disorder at high 
temperature where entropy smoothens such plateaus.
%Surprisingly, at force $ F $ greater than 0.75, we see the 
%opposite trend. This opens a new domain of the research field  
%reflecting that response of a force on percolative studies 
%may intricate interesting feature which have direct relevance
%in biology.

We have modeled the unfolding process in a cell where the
polymer is surrounded by non-interacting  biomolecules. Our results
based on exact enumeration technique clearly show that disorder induces
intermediate states because of the heterogeneity in the interaction.
The occurrence of peaks of equal height in the probability distribution
shows the coexistence of two states at different forces  suggesting that 
the transition is discontinuous in the case of finite chains. This may be 
because the underlying structure (percolation clusters)
is no longer a regular lattice but a fractal \cite{janke}.
At this stage long chain simulation is required
to verify this. It is important to recall that  experimentally 
observed coil-globule transition is also first order. 
In constant distance ensemble  system probes 
local ground state and average of force is calculated from them. 
Therefore, heterogeneity in non-bonded nearest neighbor interaction 
induced by disorder shows such strong oscillation 
in the F-x curve. At this stage  additional work is 
needed to understand the effect of a force in the cellular environment.

\section{acknowledgments}

We thank D. Dhar, S. M. Bhattacharjee and K. P. N. Murthy for 
many helpful discussions related to disorder averaging. 
We also thank I. Jensen for critical reading of this paper
and his comments.  Financial supports from DST New Delhi 
and UGC, New Delhi are gratefully acknowledged.
We also acknowledge the generous computers support from MPIPKS, 
Dresden, Germany.

\end{document}